\documentclass[12pt, a4paper, oneside]{article}

\usepackage[latin1]{inputenc}

\usepackage{latexsym,amsmath,amsthm,amsfonts,amssymb}
\usepackage{enumerate,array}
\usepackage{longtable,verbatim,wrapfig}
\usepackage{booktabs,colortbl,setspace}
\usepackage{cancel,anysize}
\usepackage{graphicx}
\usepackage{here}
\usepackage{url}
\usepackage{ragged2e}
\usepackage{multicol}

\usepackage[left=2.5cm,top=2cm,right=2.5cm,bottom=2.5cm]{geometry} 
\usepackage{geometry} 
\usepackage{anysize}
\usepackage{epsfig,graphicx,enumitem}
\usepackage{float}
\usepackage[footnotesize]{caption}
\newcommand{\bX}{\boldsymbol{\mathrm{X}}}

\newcommand{\bth}{\boldsymbol{\theta}}

\usepackage[absolute]{textpos}

\usepackage{authblk}
\title{Bayesian analysis for pretest-posttest binary outcomes with adaptive  significance levels}
\author[1]{Alejandra Estefan\'ia Pati\~no Hoyos}

\author[2]{Johnatan Cardona Jim\'enez}

\affil[1]{Instituci\'on Universitaria Pascual Bravo, Colombia; alejandra.patino@pascualbravo.edu.co}
\affil[2]{Universidad Nacional de Colombia, Colombia; jcardonj@unal.edu.co}

\date{}                     
\begin{document}

\maketitle

\begin{abstract}

Count outcomes in longitudinal studies are frequent in clinical and engineering studies. In frequentist and Bayesian statistical analysis, methods such as Mixed linear models allow the variability or correlation within individuals to be taken into account. However, in more straightforward scenarios, where only two stages of an experiment are observed (pre-treatment vs. post-treatment), there are only a few tools available, mainly for continuous outcomes. Thus, this work introduces a Bayesian statistical methodology for comparing paired samples in binary pretest-posttest scenarios. We establish a Bayesian probabilistic model for the inferential analysis of the unknown quantities, which is validated and refined through simulation analyses, and present an application to a dataset taken from the Television School and Family Smoking Prevention and Cessation Project (TVSFP) (Flay et al., 1995). The application of the Full Bayesian Significance Test (FBST) for precise hypothesis testing, along with the implementation of adaptive significance levels in the decision-making process, is included.
\end{abstract}

\section{Introduction}
In pre-post studies, the effect of an intervention is assessed by comparing a subject's outcomes before (pre) and after (post) receiving the intervention. In all these stages, each unit of analysis yields two outcomes from two different conditions. In this work, the primary interest lies in the difference between proportions when the observed variable is a binomial outcome at each stage. Since these measurements are correlated, statistical methods to compare independent samples cannot be applied. For continuous outcomes, the paired t-test is the standard statistical method in the frequentist approach for evaluating differences between means. However, the paired t-test does not apply to non-continuous variables such as binary and count outcomes. This work presents a Bayesian statistical method for comparing paired samples in binary pretest-posttest scenarios. The methodology includes the use of the Full Bayesian Significance Test (FBST) (Pereira and Stern, 1999) for precise hypothesis testing and adaptive significance levels in the decision-making process.

\section{Methodology}

\begin{itemize}
\justifying
\item Let $y_{ij}$, for $i = 1,..., n_j$, be binary random variables related to the $i$-th individual or sample unit at a time $j$, with $j = 1,2$.
\item In that way, one can define the following new random variables: $X_{j}=\sum \limits_i y_{ij}$, where $X_{j}\vert \theta_j \sim \text{bin} (n_j,\theta_j)$ with $\theta_j$ being the success probability at the stage $j$. 
 

\item  Given $\boldsymbol{\theta}=(\theta_1,\theta_2)$, $X_{1}$ and $X_{2}$ are conditionally independent, thus, their joint distribution is such that $(X_{1}, X_{2}) \vert \boldsymbol{\theta} \sim \text{bin} (n_1,\theta_1) \times \text{bin} (n_2,\theta_2)$. 

\item  The unconditional correlation between $X_{1}$ and $X_{2}$ is introduced through the bivariate distribution of $(\theta_1,\theta_2)$.
\end{itemize}

We propose as a prior distribution the three-parameter bivariate beta model proposed by Olkin and Liu (2003), also called the \textbf{two-dimensional beta distribution}, which allowed for positive correlation. That distribution has a density given by  

\begin{small}
\begin{equation}\label{prior}
f(\theta_1,\theta_2)=\dfrac{\Gamma(\alpha)}{\Gamma(\alpha_1)\Gamma(\alpha_2)\Gamma(\alpha_0)} \, \dfrac{\theta_1^{\alpha_1-1} (1-\theta_1)^{\alpha_2+\alpha_0-1}\, \theta_2^{\alpha_2-1} (1-\theta_2)^{\alpha_1+\alpha_0-1}}{(1-\theta_1\theta_2)^{\alpha}}
\end{equation}   
\end{small}

\vspace{1em}

for $0 < \theta_1 < 1$, $0 < \theta_2 < 1$, and $\alpha = \alpha_1 + \alpha_2 + \alpha_0$.

\vspace{1em}

The unconditional distribution of $(X_{1}, X_{2})$ called the \textbf{two-dimensional beta binomial distribution} is introduced by Bibby and V{\ae}th (2011) and has a probability mass function given by
\begin{small}
\begin{align}\label{margx}
f(x_{1}, x_{2})&=\int_{
[1,0]^2}f_{\bX_j \vert \boldsymbol{\theta}}(\boldsymbol{x_j} \vert \boldsymbol{\theta} )\, f(\boldsymbol{\theta})\, d\boldsymbol{\theta}\nonumber\\[3pt]
&= {n_1 \choose x_{1}} {n_2 \choose x_{2}} \dfrac{\Gamma(\alpha)}{\Gamma(\alpha_0)\Gamma(\alpha_1)\Gamma(\alpha_2)} \, \dfrac{\Gamma(x_{1}+\alpha_1)\,  \Gamma(n_1-x_{1}+\alpha-\alpha_1)}{\Gamma(n_1
+\alpha)} \nonumber\\[3pt]
&\times \dfrac{\Gamma(x_{2}+\alpha_2) \,  \Gamma(n_2-x_{2}+\alpha-\alpha_2)}{\Gamma(n_2
+\alpha)} \nonumber\\
&\times {}_3 F_2(\alpha,x_{1}+\alpha_1, x_{2}+\alpha_2; n_1+\alpha, n_2+\alpha;1)
\end{align}
\end{small}

for $x_{1}=0,1, \dots n_1$ and $x_{2}=0,1, \dots n_2$.\newline

\vspace{0.5mm}

Thus, the Bayesian posterior distribution of $\bth$ is given by

\begin{small}  
\begin{align}\label{posti}
f(\theta_1, \theta_2 \vert x_{1}, x_{2})&=
\dfrac{\Gamma(n_1+\alpha)}{\Gamma(x_{1}+\alpha_1)\Gamma(n_1-x_{1}+\alpha-\alpha_1)} \, \dfrac{\Gamma(n_2+\alpha)}{\Gamma(x_{2}+\alpha_2)\Gamma(n_2-x_{2}+\alpha-\alpha_2)} \nonumber\\[3.5pt]
&\times \dfrac{\theta_1^{\alpha_1+x_{1}-1} (1-\theta_1)^{\alpha_2+\alpha_0+n_1-x_{1}-1} \, \theta_2^{\alpha_2+x_{2}-1} (1-\theta_2)^{\alpha_1+\alpha_0+n_2-x_{2}-1}}{(1-\theta_1\theta_2)^{\alpha}}\nonumber\\[3pt]
&\times  \left[{}_3 F_2(\alpha,x_{1}+\alpha_1, x_{2}+\alpha_2; n_1+\alpha, n_2+\alpha;1)\right]^{-1}
\end{align}
\end{small}

\subsection{Full Bayesian Significance Test (FBST)}

The Full Bayesian Significance Test (FBST) for precise hypotheses is introduced by  Pereira and Stern (1999) as a Bayesian alternative to the traditional significance tests based on \textit{p}-values. With the FBST, the authors introduce the \textit{e}-value as an evidence index in favor of the null hypothesis (\textbf{H}). 

\vspace{1mm}

Consider a null hypothesis ${\text{\textbf{H}}: \boldsymbol{\theta} \in \boldsymbol{\Theta}_\text{\textbf{H}}}$. The tangential set to $\text{\textbf{H}}$ is given by
\begin{equation*}
T_{x_{10},x_{20}}=\left\lbrace (\theta_1,\theta_2)\in \boldsymbol{\Theta}: f(\theta_1,\theta_2\vert x_{10},x_{20})>\underset{\text{\textbf{H}}}{\operatorname{sup}}f(\theta_1,\theta_2\vert x_{10},x_{20}) \right\rbrace.
\end{equation*}
The measure of evidence ($e$-value) in favor of $\text{\textbf{H}}$ is defined as the $T_{x_{10},x_{20}}$ posterior probability complement, that is, 
\begin{align*}
ev\left(\text{\textbf{H}};x_{10},x_{20}\right)&=1-P((\theta_1,\theta_2) \in T_{x_{10},x_{20}}\vert x_0)\\
&=1- \int_{} \int_{T_{x_{10},x_{20}}}f(\theta_1,\theta_2\vert x_{10},x_{20})\, d\theta_1 d\theta_2.  
\end{align*}
The \textbf{FBST} is the procedure that rejects $\text{\textbf{H}}$ whenever $ev\left(\text{\textbf{H}};x_{10},x_{20}\right)$ is small (Pereira et al. , 2008).

\subsection{Adaptive cut-off values for evidence in the FBST}

Consider the test
\vspace{-5pt}
\begin{equation*}
\varphi_{e}(x_{1},x_{2})= \left\{ \begin{array}{l}  0 \quad if \quad ev\left(\text{\textbf{H}};x_{1},x_{2}\right)> k\\ \\ 1 \quad if \quad ev\left(\text{\textbf{H}};x_{1},x_{2}\right)\leq 
k.  \end{array} \right.\;
\end{equation*}
\vspace{-2pt}

Thus, the power function and the averaged error probabilities are given by
\begin{small}
\begin{align*}
\pi_{\varphi_{e}}(\boldsymbol{\theta})&=P\left(\lbrace (X_{1}, X_{2}) \in \Omega:\varphi_{e}(X_{1}, X_{2})=1 \rbrace\vert \,\boldsymbol{\theta}\right)\\
&=P\left( \left\lbrace (X_{1}, X_{2}) \in \Omega: ev\left(\text{\textbf{H}};X_{1}, X_{2}\right)\leq 
k\right\rbrace \Bigm| \boldsymbol{\theta}\right),
\end{align*}
\end{small}
\vspace{-10pt}
\begin{small}
\begin{equation*}\label{erro1}
\alpha_{\varphi_{e}}= E_{\boldsymbol{\theta}}\left[ \pi_{\varphi_{e}}(\boldsymbol{\theta})\vert \text{\textbf{H}}\right], \quad \text{and} \quad \beta_{\varphi_{e}}=E_{\boldsymbol{\theta}}\left[ 1-\pi_{\varphi_{e}}(\boldsymbol{\theta})\vert \text{\textbf{A}}\right].
\end{equation*}
\end{small}

The adaptive cut-off value $k^{*}$ for $ev\left(\text{\textbf{H}};x\right)$ will be the $k$ that minimizes $a\alpha_{\varphi_{e}}+b\beta_{\varphi_{e}}$ (Pati\~no, et al., 2023).

\section{Simulation study}
\subsection{Non-informative prior}

  \begin{figure}[H]
\setlength{\tabcolsep}{18pt}
\begin{tabular}{cc}
    \includegraphics[scale=0.5]{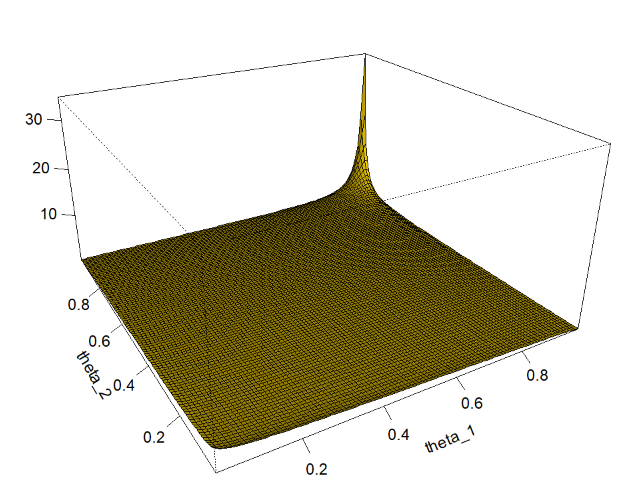} &\includegraphics[scale=0.5]{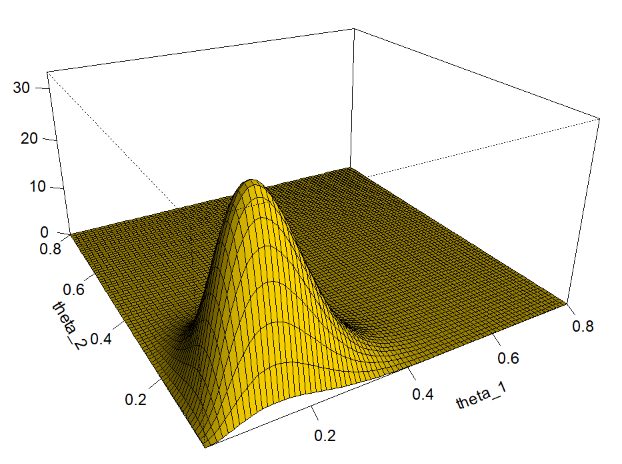} \\
     \scriptsize (a) &  \scriptsize (b) 
 \end{tabular}

\caption{\footnotesize (a) Non-informative prior approximated by Kullback-Liebler divergence. (b) Posterior distribution of $\bth$, taking  $\bX_i \vert \boldsymbol{\theta} \sim \text{bin} (n_1,\theta_1) \times \text{bin} (n_2,\theta_2)$ with $n_1=n_2=20$ and $\theta_1=\theta_2=0.1$.}
\end{figure}

\vspace{0.5cm}

\begin{figure}[H]
\setlength{\tabcolsep}{15pt}
\begin{tabular}{cc}
    \includegraphics[scale=0.4]{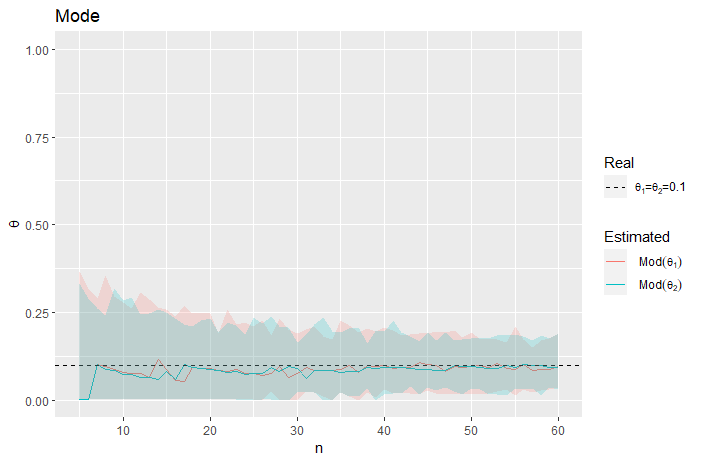} &\includegraphics[scale=0.4]{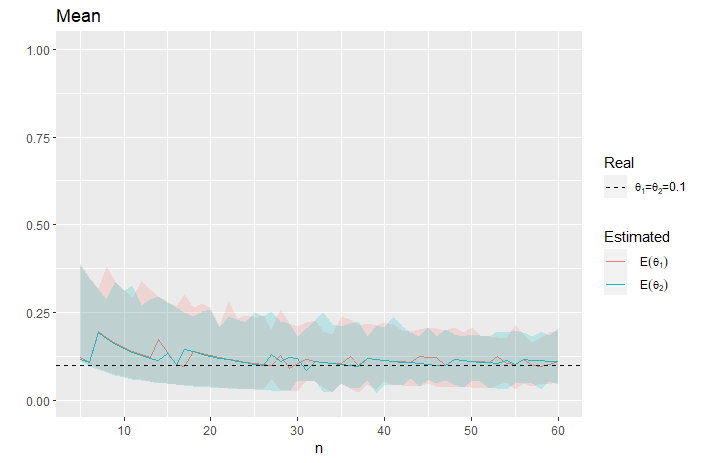} \\
     \scriptsize (a) &  \scriptsize (b) 
 \end{tabular}

\caption{\footnotesize MCMC estimation of $\bth$ by the mode (a)  and the mean (b) of the posterior distribution as a function of $n$, taking  $\bX_i \vert \boldsymbol{\theta} \sim \text{bin} (n_1,\theta_1) \times \text{bin} (n_2,\theta_2)$ with $\theta_1=\theta_2=0.1$.}
\end{figure}

\subsection{Informative Prior-Data Conflict}

\begin{figure}[H]
\setlength{\tabcolsep}{18pt}
\begin{tabular}{cc}
    \includegraphics[scale=0.5]{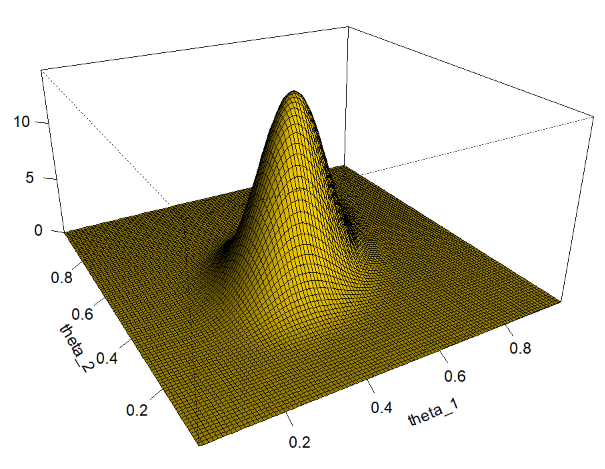} &\includegraphics[scale=0.5]{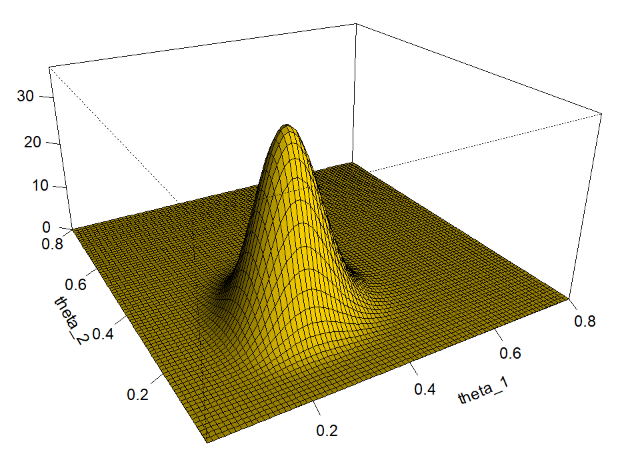} \\
     \scriptsize (a) &  \scriptsize (b) 
 \end{tabular}

\caption{\footnotesize (a) Informative prior with $\alpha_1 = \alpha_2 = \alpha_0=10$. (b) Posterior distribution of $\bth$, taking  $\bX_i \vert \boldsymbol{\theta} \sim \text{bin} (n_1,\theta_1) \times \text{bin} (n_2,\theta_2)$ with $n_1=n_2=20$ and $\theta_1=\theta_2=0.1$.}
\end{figure}

\vspace{0.5cm}

\begin{figure}[H]
\setlength{\tabcolsep}{15pt}
\begin{tabular}{cc}
    \includegraphics[scale=0.4]{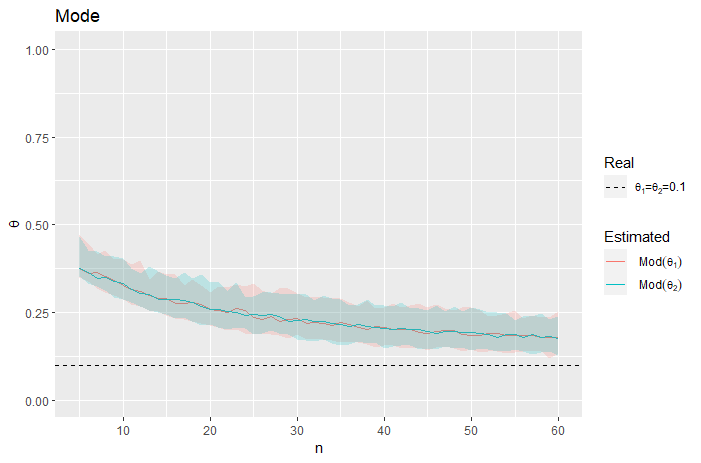} &\includegraphics[scale=0.4]{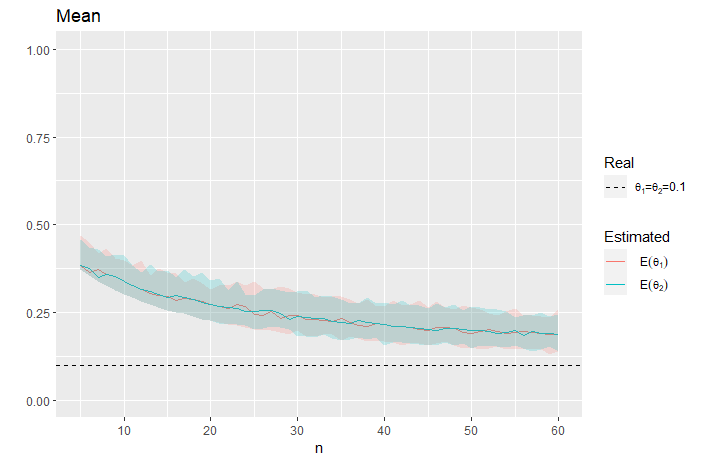} \\
     \scriptsize (a) &  \scriptsize (b) 
 \end{tabular}

\caption{\footnotesize MCMC estimation of $\bth$ by the mode (a)  and the mean (b) of the posterior distribution as a function of $n$, taking  $\bX_i \vert \boldsymbol{\theta} \sim \text{bin} (n_1,\theta_1) \times \text{bin} (n_2,\theta_2)$ with $\theta_1=\theta_2=0.1$.}
\end{figure}

\subsection{Informative Prior-Data No Conflict}

\begin{figure}[H]
\setlength{\tabcolsep}{18pt}
\begin{tabular}{cc}
    \includegraphics[scale=0.5]{Figures/prior_inf_Confl.png} &\includegraphics[scale=0.5]{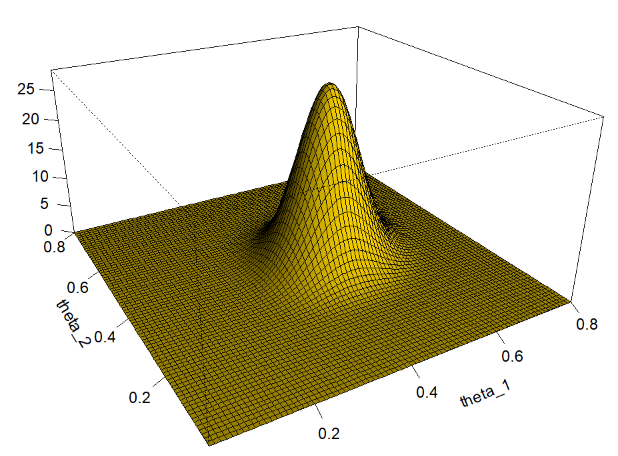} \\
     \scriptsize (a) &  \scriptsize (b) 
 \end{tabular}

\caption{\footnotesize (a) Informative prior with $\alpha_1 = \alpha_2 = \alpha_0=10$. (b) Posterior distribution of $\bth$, taking  $\bX_i \vert \boldsymbol{\theta} \sim \text{bin} (n_1,\theta_1) \times \text{bin} (n_2,\theta_2)$ with $n_1=n_2=20$ and $\theta_1=\theta_2=0.5$.}
\end{figure}

\vspace{0.5cm}

\begin{figure}[H]
\setlength{\tabcolsep}{15pt}
\begin{tabular}{cc}
    \includegraphics[scale=0.4]{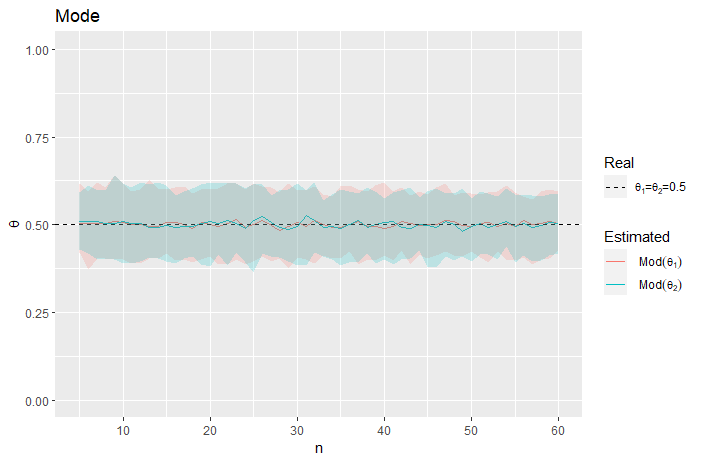} &\includegraphics[scale=0.4]{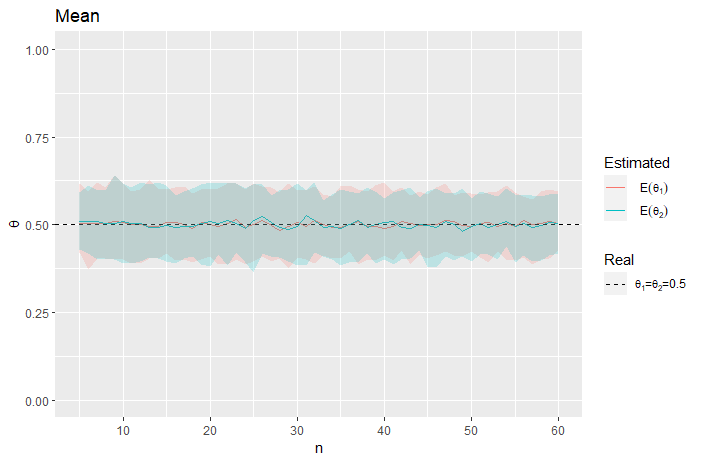} \\
     \scriptsize (a) &  \scriptsize (b) 
 \end{tabular}

\caption{\footnotesize MCMC estimation of $\bth$ by the mode (a)  and the mean (b) of the posterior distribution as a function of $n$, taking  $\bX_i \vert \boldsymbol{\theta} \sim \text{bin} (n_1,\theta_1) \times \text{bin} (n_2,\theta_2)$ with $\theta_1=\theta_2=0.5$.}
\end{figure}

 \section{Application}
 
The data used in this application is taken from the Television School and Family Smoking Prevention and Cessation Project (TVSFP) (Flay et al., 1995), which investigated the effectiveness of a school-based social-resistance classroom curriculum (CC) and a television-based program (TV) for tobacco use prevention and cessation. The primary outcome measure is the tobacco and health knowledge scale (THKS) score, which ranges from zero to seven. For analysis, scores are dichotomized into 0-2 and 3-7, following the approach of Hedeker (1999).





 \vspace*{2mm}

A partial list of these data is given in Table 1:
\begin{table}[H]
\small
\centering
\setlength{\tabcolsep}{3.5pt}
\bgroup
\def\arraystretch{0.9}
\begin{tabular}{cccccccc}
  \hline \hline
\textbf{id}	&	\textbf{school}	&	\textbf{class}	&	\textbf{school.based}	&	\textbf{tv.based}	&	\textbf{stage}	&	\textbf{THKS}	&	\textbf{binTHKS}
\\ 
  \hline
24	&	404	&	404101	&	yes	&	yes	&	pre	&	1	&	0\\
24	&	404	&	404101	&	yes	&	yes	&	post	&	3	&	1\\
25	&	404	&	404101	&	yes	&	yes	&	pre	&	2	&	0\\
25	&	404	&	404101	&	yes	&	yes	&	post	&	4	&	1\\
26	&	404	&	404101	&	yes	&	yes	&	pre	&	4	&	1\\
26	&	404	&	404101	&	yes	&	yes	&	post	&	2	&	0\\
   \hline    
   \hline 
\end{tabular}
 \egroup

\caption{ \footnotesize TVSFP Project partial dataset.}
\end{table}

The hypotheses that could be calculated are given by
\vspace{2pt}
\begin{itemize}

\item \textbf{Non-precise hypotheses:}
\begin{small}
\begin{equation*}
\text{\textbf{H}}:\theta_1  \leq \theta_{2} \,\,\,\,\,\, vs. \,\,\,\,\,\,\text{\textbf{A}}: \theta_1  > \theta_{2}
\end{equation*}
\end{small}
\item \textbf{Precise hypotheses:}
\begin{small}
\begin{equation*}
\text{\textbf{H}}:\theta_1  = \theta_{2} \,\,\,\,\,\, vs. \,\,\,\,\,\,\text{\textbf{A}}: \theta_1  \neq \theta_{2}
\end{equation*}
\end{small}
\end{itemize}

\begin{figure}[H]
\setlength{\tabcolsep}{18pt}
\begin{tabular}{cc}
    \includegraphics[scale=0.35]{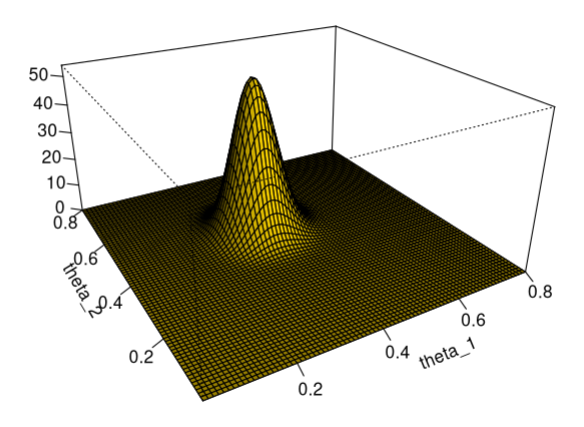} &\includegraphics[scale=0.4]{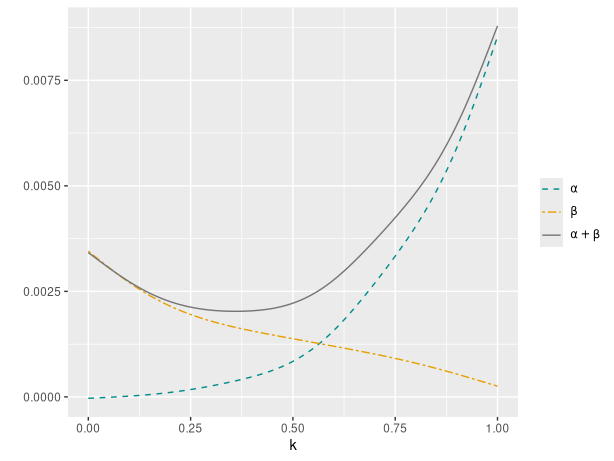} \\
\scriptsize (a)   & \scriptsize(b) 
 \end{tabular}

\caption{\footnotesize (a) Posterior distribution with non-informative prior approximated by Kullback-Liebler divergence  (b) Averaged error probabilities ($\alpha_{\varphi_e}$, $\beta_{\varphi_e}$ and $\alpha_{\varphi_e}+\beta_{\varphi_e}$) as function of $k$. Sample size $n_1=n_2=164$,  $a=b=1$, $M=500$. (in no intervention (No CC - No TV))}
\end{figure}

\begin{table}[H]
\centering
\setlength{\tabcolsep}{3.5pt}
\begin{footnotesize}
\renewcommand{\arraystretch}{1.5} 
\setlength\extrarowheight{-3pt}
\begin{tabular}{ccccc}
  \hline\hline
  & \textbf{Non-precise} & \textbf{Precise}\\ 
\cline{2-4} 
 \textbf{Study conditions}  & $P(\text{\textbf{H}}\vert \bX)$ & $ev\left(\text{\textbf{H}};\bX\right)$ & $k^*$& $n_1=n_2$\\ 
  \hline 
\textbf{CC - TV} & 0.9999 & 0.0005&0.4515& 50  \\ 
 \textbf{No CC - TV} & 0.4961 & 1& 0.3153&164 \\ 
 \textbf{CC - No TV} & 0.9966 & 0.0184& 0.4124&76 \\ 
  \textbf{No CC - No TV} & 0.9372 & 0.2874& 0.3604&160\\ 
 
     \hline\hline
\end{tabular}
\end{footnotesize}
\caption{\footnotesize TVSFP Project: Bayesian hypothesis testing.}
\label{tabknpriorisNN}
\end{table}

\section*{References}

\begin{description}

{\item[] Bibby, B. M. and Michael V{\ae}th, M. (2011). The
two-dimensional beta binomial distribution. {{ \it Statistics \& Probability
Letters}, {\bf 81} (7):884-891.}}

{\item[] Flay, B. R. et al. (1995). The television, school, and family smoking prevention and
cessation project. viii. student outcomes and mediating variables. {{\it Prev
Med}, {\bf 24} (1).}}

{\item[] Hedeker, D. (1999). Mixno: a computer program for mixed-effects nominal logistic regression. {{\it Journal of Statistical Software}, {\bf 4(5)}:1-92.}}

{\item[] Olkin, I. and Liu, R. (2003). A bivariate beta
distribution. {{ \it Statistics \& Probability Letters}, {\bf 62} (4):407-412.}}

{\item[] Pati\~no, A. E., Fossaluza, V., Esteves, L. G. and Pereira, C. A. B. (2023). Adaptive significance levels in tests for linear regression models: The e-value and p-value cases. {{\it Entropy}, {\bf 25} (1).}}

{\item[] Pereira, C. A. B.and Stern, J. M.  (1999). 
Evidence and Credibility: Full Bayesian Significance Test for Precise Hypotheses. {{\it Entropy}, {\bf 4}, 99-110.}}


\end{description}

\end{document}